\documentclass[modern]{aastex701}

\usepackage{amsmath}
\usepackage{wasysym}
\newcommand{\pdf}{\mathrm{Pr}}

\begin{document}

\title{What to make of the Earth's curiously intermediate land fraction?}

\author[orcid=0000-0002-4365-7366,sname='Kipping']{David Kipping}
\affiliation{Columbia University, 550 W 120th Street, New York NY 10027}
\email[show]{dkipping@astro.columbia.edu}  

\begin{abstract}
Approximately two-thirds of the Earth, the only known inhabited planet, is covered in ocean. Why not 0.01\% or 99.99\%? It has been previously suggested that this may represent a certain degree of fine-tuning, and thus perhaps observers are a-priori more likely to develop on those rare worlds with nearly equal land-ocean ratios, such as our own. In this work, we take the single datum of the Earth and then use Bayesian inference to compare four models for the probability distribution of a planet becoming inhabited by observers as a function of land-fraction, $f$, which we classify as i) land-centric ii) ocean-centric iii) equi-centric and iv) indifference. We find that no model is \textit{strongly} favoured over the others, but that 1) the land-centric model is disfavoured over all others, and, 2) the equi-centric model is favoured over all competitors. Further, we show that more extreme models with heavy tail-weighting are strongly disfavoured even when conditioned upon the Earth alone. For example, a land-centric model where the median planet has $f=0.82$ (or greater) is in strong tension with our existence. Finally, we consider the potential addition of more data via Mars or exoplanets. Should paleo-Mars have once harboured life and had $f<0.20$, then this would strongly favour the ocean-centric model for life, over a land-centric hypothesis. We show that strong evidence for/against the equi-centric model versus its competitors would likely require at least a dozen inhabited exoplanets, offering a well-motivated sample size for future experiments.
\end{abstract}



\section{Introduction}
\label{sec:intro}

The Earth is our sole example of what an inhabited planet looks like. We therefore know with certainty that a planet such as the Earth is capable of maintaining not just life, but a rich biosphere brimming with complex life and observers such as ourselves. For this reason, astronomers have often framed the search for Earth analogs of being of central importance in our search for life amongst the stars \citep{tpf:2002,kepler:2010,luvoir:2019,habex:2020}. Of course, astronomers and planetary scientists certainly entertain the prospect of life in other environments too, such as icy moons \citep{kimura:2015}, hycean worlds \citep{madhu:2024}, Titan \citep{lunine:2009} and minor bodies \citep{wordsworth:2024} - but unlike Earth-like bodies, this remains speculation.

Many features of the Earth have been argued to be atypical of even terrestrial planets within the habitable zone, such as a large Moon, a magnetic field and plate tectonics forming the so-called ``Rare Earth'' hypothesis \citep{ward:2000}. In addition to these, it has been argued that the fraction of the Earth covered by water, about two-thirds, has a suspiciously intermediate value. Notably, \citet{simpson:2017} argue that the water basin capacity of the Earth is sufficiently deep that an intermediate land fraction, such as we observe, appears \textit{a-priori} unlikely. Nevertheless, considerable uncertainty remains about Earth's deep water cycle. For example, estimates of the amount of water in the mantle vary from one quarter to four times that on the surface \citep{hirschmann:2006}.

Due to our evolving and uncertain understanding of even water within the Earth, we argue that it is instructive to consider a more agnostic approach to what the distribution of the land-fraction, $f$, might be. A key question we consider in this work is whether there is any evidence for a Rare Earth type argument here - are observers more likely to emerge on planets with intermediate land fractions? We consider that the only bullet-proof evidence in hand is that we exist on a planet with a land fraction\footnote{
It should be noted that $f$ evolves over geological timescales, and was likely lower in the past \citep{johnson:2020}. However, this study is focussed on the emergence of intelligent observers (a recent phenomenon) and thus these previous epochs don't provide additional input data to this work.
} of $f=0.292$ - we eject any assumptions about water basin capacity, evolutionary preferences or otherwise. This stripped down analysis sacrifices precision - fewer assumptions means looser constraints. But, it excels in accuracy - fewer assumptions means a more robust result.

In Section~\ref{sec:simplemodels}, we present four simple candidate models for the observer selection effect - what types of worlds favour the emergence of observers. We then compare the models in a Bayesian model comparison framework. In Section~\ref{sec:extrememodels}, we consider more extreme observer selection effect models that are strongly disfavoured by our single datum - us. We consider the impact of future additional data in Section~\ref{sec:mars}, before providing a discussion and detailed comparison to the work of \citet{simpson:2017} in Section~\ref{sec:discussion}.

\section{Comparing Simple Models}
\label{sec:simplemodels}

\subsection{Deriving Posterior Probabilities}
\label{sub:posteriors}

\subsubsection{Prior on $f$}
\label{sub2:fprior}

The first question we must address to make progress is: what is the \textit{a-priori} probability distribution for the land-fraction, $f$, amongst rocky planets in the temperate zone of their stars? Let us dub this function, $\pdf(f)$. It is argued in this work that we possess no prior information concerning this point at the time of writing and thus we seek a so-called reference/non-informative prior. Given that $f$ is bound over the interval $[0,1]$, the corresponding Jeffrey's prior is the Jaynes' prior \citep{jaynes:1968}, given by

\begin{align}
\pdf(f) &= \frac{1}{\pi \sqrt{f(1-f)}}.
\label{eqn:fprior}
\end{align}

The Jaynes prior places the greatest weight at the boundaries ($f=0$ and $f=1$) and the lowest (but still substantial) weight at the intermediate value of $f=\tfrac{1}{2}$. It thus follows Haldane's perspective that finely-tuned intermediate solutions are less likely (\textit{a-priori}) that saturated configuration \citep{haldane:1932}. This is perhaps most famously illustrated by the thought experiment of considering trying to dissolve an unknown chemical into a series of water beakers (with slightly different temperatures, acidity, etc). One should reasonably expect the unknown chemical will dissolve in approximately all of the beakers or essentially none - but a 50:50 outcome would be extraordinarily surprising.

\subsubsection{Model L: The Land-Centric Likelihood}
\label{sub2:modelL}

The next step is to define a likelihood function and thus it is first necessary to define the likelihood of \textit{what} exactly? Following \citet{kipping:2020}, we argue that an implicit datum in hand is that we exist, ``cogito ergo sum'' (CES). Thus, we can ask, what is the probability of CES given a certain choice of $f$ - which represents a likelihood. Many possible functions could be suggested at this stage for such a likelihood. Rather than adopt any one model ad-hoc, we consider four variants and then weigh them against one another using Bayesian model comparison.

The first model we consider is that $\pdf(\mathrm{CES}|f) \propto f$. In other words, observers such as ourselves are more likely to emerge on planets with large land fractions. We resist the urge to speculate about the mechanism behind such a bias in this work - we merely aim to compare each possibility fairly. This can be formalised as

\begin{align}
\pdf(\mathrm{CES}|f,\mathcal{M}_L) &\propto f.
\label{eqn:likelihoodL}
\end{align}

Note how we add the notation $\mathcal{M}_L$, the land-centric model, explicitly into the conditional. The posterior is now

\begin{align}
\pdf(f|\mathrm{CES},\mathcal{M}_L) &= \frac{2 f}{\pi \sqrt{f(1-f)}},
\label{eqn:posteriorL}
\end{align}

where we have added the appropriate normalization to ensure $
\int \pdf(f|\mathrm{CES},\mathcal{M}_L)\,\mathrm{d}f = 1$.

\subsubsection{Model O: The Ocean-Centric Likelihood}
\label{sub2:modelO}

The natural competing model to this is an ocean-centric likelihood, such that $\pdf(\mathrm{CES}|f) \propto (1-f)$. Following the same steps as before, one obtains

\begin{align}
\pdf(f|\mathrm{CES},\mathcal{M}_O) &= \frac{2 (1-f)}{\pi \sqrt{f(1-f)}}.
\label{eqn:posteriorO}
\end{align}

\subsubsection{Model E: The Equi-Centric Likelihood}
\label{sub2:modelE}

A third model we consider is one that favours planets with an equal mix of land and ocean, i.e. $f\simeq\tfrac{1}{2}$. Such a model was explicitly proposed in the recent study of \citet{stern:2024}. Since the previous two models both had zero weight at one of their boundaries, we seek an equi-centric likelihood that has zero weight at both boundaries. We require it to peak at $f=\tfrac{1}{2}$ and one might naturally consider some kind of Gaussian-like shape. To this end, we found that the Beta distribution with $\alpha=\beta=2$ does a good job of capturing these requirements:

\begin{align}
\pdf(\mathrm{CES}|f,\mathcal{M}_E) &\propto f (1-f).
\label{eqn:likelihoodE}
\end{align}

The corresponding posterior may be found to be:

\begin{align}
\pdf(f|\mathrm{CES},\mathcal{M}_E) &= \frac{8 \sqrt{f(1-f)}}{\pi}.
\label{eqn:posteriorE}
\end{align}

\subsubsection{Model I: The Indifferent Likelihood}
\label{sub2:modelI}

Finally, one may state an alternative model is that the probability of emerging on a planet is completely indifferent to the land fraction, $f$. Formally, one might thus invoke a uniform likelihood function, but the end result is of course that the prior equals the posterior.

\begin{align}
\pdf(f|\mathrm{CES},\mathcal{M}_I) &= \frac{1}{\pi \sqrt{f(1-f)}}.
\label{eqn:posteriorI}
\end{align}

\begin{figure}
\begin{center}
\includegraphics[width=15.5 cm]{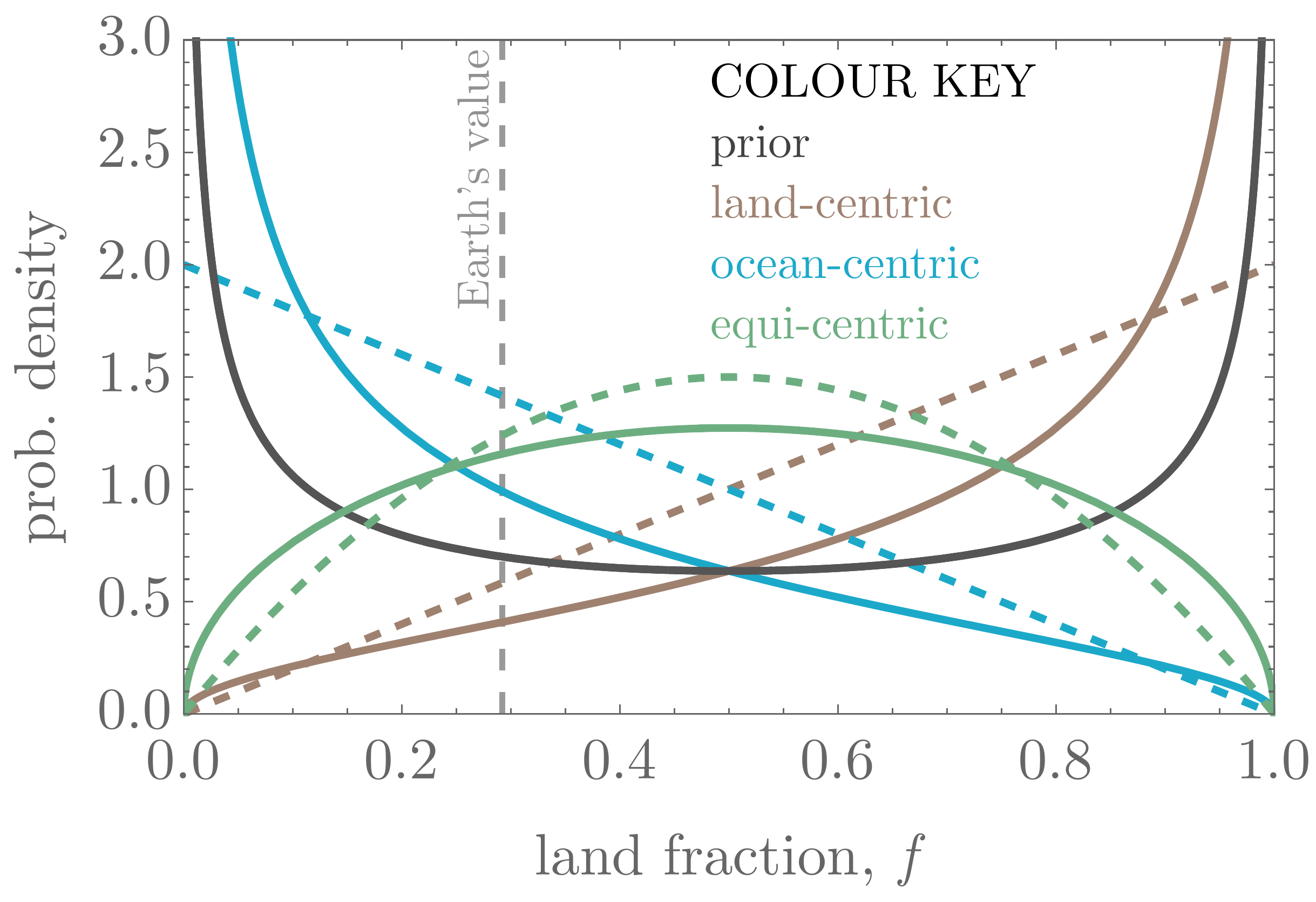}
\caption{\emph{
Comparison of the posteriors (solid lines) from our four different fiducial models, labeled inset. The dashed-lines represent the likelihood functions of each, and the solid grey line is the prior - which is identical to the fourth model of indifference and thus not explicitly shown.
}} 
\label{fig:posteriors}
\end{center}
\end{figure}

The four different posteriors are compared in Figure~\ref{fig:posteriors}, where we note that the indifference model's posterior equals the prior and thus is not explicitly shown.

\subsection{Bayes Factors}
\label{sub:bayesfactors}

We are now ready to ingest the datum that the Earth has $f_{\oplus}=0.292$ (marked by the vertical line on Figure~\ref{fig:posteriors} and weigh the models against one another. By construction, one may do this may simply calculating the ratio of their likelihoods evaluated at the datum $f=f_{\oplus}$. So, for example, the Bayes factor of model 2 versus model 1 will be

\begin{align}
B_{21} &= \frac{ \pdf(f=f_{\oplus}|\mathrm{CES},\mathcal{M}_2) }{ \pdf(f=f_{\oplus}|\mathrm{CES},\mathcal{M}_1) }.
\label{eqn:B12}
\end{align}

We calculated all possible pairs of Bayes factors in the upper-panel of Table~\ref{tab:n1_matrix}. As that table shows, none of the Bayes factors exceed $10$ nor fall below $0.1$. In other words, there is no strong evidence for any model over any other model \citep{jeffreys:1946,kass:1995}. The largest Bayes factor occurs between the ocean- and land-centric models, yielding $B_{LO} = 2.45$. In other words, the ocean-centric hypothesis is approximately two-and-a-half times more likely than the land-centric hypothesis given Earth's land fraction, if we assume no \textit{a-priori} preference between them.

\begin{table}
\caption{
Matrix table comparing the Bayes factors between various pairs of models, conditioned upon the fact Earth has a land fraction of $f_{\oplus}=0.292$. The column header represents the numerator and the row header represents the denominator. For example, in the top-panel, the value of $2.42$ corresponds to $B_{OL}$, the Bayes factor of the ocean-centric (O) model over the land-centric (L) model. The upper-panel assumes a Jaynes prior, the lower-panel assumes a uniform prior in $f$.
} 
\centering 
\begin{tabular}{c c c c c c} 
 & \vline & $\mathcal{M}_I$ & $\mathcal{M}_E$ & $\mathcal{M}_L$ & $\mathcal{M}_O$ \\ [0.5ex] 
\hline
$\mathcal{M}_I$ & \vline & 1 	& 1.65	& 0.58	& 1.41	\\
$\mathcal{M}_E$ & \vline & 0.60 	& 1		& 0.35	& 0.86	\\
$\mathcal{M}_L$ & \vline & 1.71 	& 2.83	& 1		& 2.42	\\
$\mathcal{M}_O$ & \vline & 0.71 	& 1.17	& 0.41	& 1		\\ [0.5ex]
\hline 
$\mathcal{M}_I$ & \vline & 1 	& 1.24	& 0.58	& 1.42	\\
$\mathcal{M}_E$ & \vline & 0.81 	& 1		& 0.47	& 1.14	\\
$\mathcal{M}_L$ & \vline & 1.71 	& 2.12	& 1		& 2.42	\\
$\mathcal{M}_O$ & \vline & 0.71 	& 0.88	& 0.41	& 1		\\ [0.5ex]
\hline
\end{tabular}
\label{tab:n1_matrix} 
\end{table}

For completion, we repeated the numerical calculations with a uniform prior ($\pdf(f) = 1$) and obtained broadly similar results. These are presented in the lower-panel of Table~\ref{tab:n1_matrix}.

Two consistent behaviours stand out from Table~\ref{tab:n1_matrix}; results that hold irrespective of which prior is used. First, the land-centric model is \textit{never} favoured over another model - it loses every comparison, which of course makes sense given that we live on an ocean-dominated world. Second, the equi-centric model is \textit{always} favoured over other models - it wins every comparison. This is less obvious given $f_{\oplus}=0.292$, from which it's not immediately obvious whether that would favour the equi-centric model over the ocean-centric model (although we note the preference is marginal).

\subsection{Critical $f$ Values}

Although Earth's $f$ value does not cause any of the Bayes factors to exceed a fiducial threshold of $10$ (typically interpreted to denote ``strong evidence''), it is instructive to reverse-engineer what $f$ value would have such an effect. Setting $B=10$ and solving for $f$ we obtain the values presented in Table~\ref{tab:n1_fcrits}. 

\begin{table}
\caption{
Matrix table evaluating how extreme Earth's land fraction would have needed to be in order to have led to Bayes factors exceeding $10$ (``strong'' evidence). The column header represents the numerator and the row header represents the denominator.  For example, the entry of $f<0.050$ corresponds to $B_{IL}$, the Bayes factor of the indifferent (I) model over the land-centric (L) model. The upper-panel assumes a Jaynes prior, the lower-panel assumes a uniform prior in $f$. A ``-'' means no choice of $f$ within the interval $[0,1]$ provides strong evidence.
$\star$ indicates that $(1-f)$ greater than this value has the same effect.
} 
\centering 
\begin{tabular}{c c c c c c} 
 & \vline & $\mathcal{M}_I$ & $\mathcal{M}_E$ & $\mathcal{M}_L$ & $\mathcal{M}_O$ \\ [0.5ex] 
\hline
$\mathcal{M}_I$ & \vline & -					& -		& -			& -		\\
$\mathcal{M}_E$ & \vline & $f<0.013^{\star}$	& -		& $f>0.975$	& $f<0.025$	\\
$\mathcal{M}_L$ & \vline & $f<0.050$ 		& -		& -			& $f<0.091$	\\
$\mathcal{M}_O$ & \vline & $f>0.950$			& -		& $f>0.909$	& -		\\ [0.5ex]
\hline
$\mathcal{M}_I$ & \vline & -					& -		& -			& -		\\
$\mathcal{M}_E$ & \vline & $f<0.017^{\star}$	& -		& $f>0.967$	& $f<0.033$	\\
$\mathcal{M}_L$ & \vline & $f<0.050$ 		& -		& -			& $f<0.091$	\\
$\mathcal{M}_O$ & \vline & $f>0.950$			& -		& $f>0.909$	& -		\\ [0.5ex]
\hline 
\end{tabular}
\label{tab:n1_fcrits} 
\end{table}

From this table, consider again the Bayes factor of the ocean-centric hypothesis over the land-centric. In this case, we find that Earth's land fraction would need to be less than 9.1\% in order to produce strong evidence favouring the ocean-centric model here. In other words, Earth's land fraction would need to be about three times less. This demonstrates that even a single data point can be constraining, at least in principle.

\section{Extreme Models Rejected by the Earth}
\label{sec:extrememodels}

\subsection{More Extreme Models}
\label{sub:extremeposteriors}

As established in the previous section, the Earth's land-fraction of $0.292$ does not lead to strong evidence for any one of our four considered models over another. The previous section explored how extreme $f_{\oplus}$ would have to be to reach such a threshold, but we can also flip this around and ask - are there more extreme models for which $f_{\oplus}=0.292$ already rejects\footnote{By which here we mean Bayes factors exceeding $10$ and thus we use the word ``rejects'' loosely.}.

The obvious way to make our models more extreme is to add an index-power to them. For example, one could modify the land-centric likelihood function in Equation~(\ref{eqn:likelihoodLn}) to:

\begin{align}
\pdf(\mathrm{CES}|f,n,\mathcal{M}_L') &\propto \frac{f^n}{n+1}.
\label{eqn:likelihoodLn}
\end{align}

For all $n>1$, this represents an even more land-centric model. Presumably, if we make $n$ large enough, the Bayes factors from before will cross our fiducial threshold of $10$. We repeated the calculation of the posteriors with the inclusion of the indexed power for the land-, ocean- and equi-centric models (the indifferent model is insensitive to such an operation). We dub these modified models with a dash and obtain:

\begin{align}
\pdf(f|\mathrm{CES},n,\mathcal{M}_L') &= \Big(\frac{\Gamma[n+1]}{\Gamma[n+\tfrac{1}{2}]}\Big) \Big(\frac{f^n}{\sqrt{\pi f (1-f)}}\Big),\\
\pdf(f|\mathrm{CES},n,\mathcal{M}_O') &= \Big(\frac{\Gamma[n+1]}{\Gamma[n+\tfrac{1}{2}]}\Big) \Big(\frac{(1-f)^{n-1/2}}{\sqrt{\pi f}}\Big),\\
\pdf(f|\mathrm{CES},n,\mathcal{M}_E') &= \Big(\frac{\Gamma[n+1]}{\Gamma[n+\tfrac{1}{2}]}\Big) \Big(\frac{4^n(f(1-f))^{n-1/2}}{\sqrt{\pi}}\Big).
\end{align}

\subsection{Excluded Values of $n$}
\label{sub:excludedn}

Conditioned upon Earth's $f_{\oplus}=0.292$, we evaluated the Bayes factors between pairs of models, as before, but keeping $n$ symbolic. We then set the Bayes factor to $10$ and solved for the corresponding $n$, imposing the condition that $n>1$ and must be real. The $n$ values which can tentatively rejected, based on our existence upon the Earth, are given in Table~\ref{tab:ncrits}.

\begin{table}
\caption{
Matrix table evaluating how extreme our competing models need to become (characterized by the index $n$) in order for Earth's land fraction ($f_{\oplus}=0.292$) to produce Bayes factors exceeding $10$ (``strong'' evidence). The column header represents the numerator and the row header represents the denominator.
} 
\centering 
\begin{tabular}{c c c c c} 
\hline\hline
- & $\mathcal{M}_I$ & $\mathcal{M}_E$ & $\mathcal{M}_L$ & $\mathcal{M}_O$ \\ [0.5ex] 
\hline
$\mathcal{M}_I$ & -			& NP		& NP		& NP		\\
$\mathcal{M}_E$ & $n>23.46$	& -			& NP		& NP		\\
$\mathcal{M}_L$ & $n>2.79$ 	& $n>2.21$	& -			& $n>2.60$	\\
$\mathcal{M}_O$ & $n>11.95$	& $n>14.83$	& NP		& -			\\ [0.5ex]
\hline 
\end{tabular}
\label{tab:ncrits} 
\end{table}

The least extreme of these is for the Bayes factor of the equi-centric model over the land-centric model, $B_{EL}$. Here, we find that any values of $n>2.21$ leads to even the single datum of Earth's $f_{\oplus}=0.229$ placing ``strong'' evidence in favour of $\mathcal{M}_E$ over $\mathcal{M}_L$. In fact, the land-centric model doesn't fare much better when compared to the other models either. Against $\mathcal{M}_O$, any coefficient of $n>2.60$ is excluded and against $\mathcal{M}_I$, any coefficient of $n>2.79$ is excluded. Together, then these results indicate that even modestly boosted land-centric models are incompatible with Earth's existence already.

The mean of these three quoted values is 2.53 and thus is approximately $5/2$. Taking the posterior distribution $\pdf(f|\mathrm{CES},n=5/2,\mathcal{M}_L')$, we find that a fraction of $1-g^{7/2}$ of its volume lives within the range $f=[g,1]$. The 50th percentile of volume thus occurs at $g=0.82$. Accordingly, this means that our own existence upon the Earth rejects a land-centric hypothesis where the median planet has a land fraction of 82\% or greater. This doesn't exclude the possibility of Arrakis-like inhabited planets out there, but it does imply that it would be very unlikely they \textit{dominate} the population of inhabited worlds.

\section{Another Data Point}
\label{sec:mars}

\subsection{Another Earth Example}
\label{sub:anotherearth}

Thus far, our analysis is heavily limited by the fact we have just a single datum - the Earth. But what if another inhabited planet were found? Perhaps the most tangible example would be that evidence for life was found on paleo-Mars, which indeed was likely covered in bodies of water in its past. Estimates for the fraction of Mars covered in ocean are highly uncertain and range from 19\% (using deuterium measurements; \citealt{villanueva:2015}) to 
75\% (using topography; \citealt{smith:1999}). These correspond to a range of $f_{\mars}=0.25$ to $f_{\mars}=0.81$ and thus clearly it is difficult to confidently assert a canonical value for paleo-Mars.

Ingesting Mars into our analysis is complicated by the fact that even if we were to one day detect evidence for past life on the red planet, that is not the same thing as an \textit{observer}. Regardless, the addition of a hypothetical new data point is still instructive, especially if we proceed under the looser conditional of an inhabited planet, rather than an observer.

To see the influence of a second datum, we thus decided to simply set this second planet to share the same $f$ value as the Earth i.e. $f_{\mars}=f_{\oplus}=0.292$. Table~\ref{tab:mars_matrix} is a re-calculation of Table~\ref{tab:n1_matrix}, but adding in this hypothetical second datum. As expected, the Bayes factors are all amplified, but still fail to cross the ``strong'' evidence threshold.

\begin{table}
\caption{
Matrix table comparing the Bayes factors between various pairs of models, conditioned upon the hypothetical scenario that both Earth and Mars has a land fraction of $f_{\oplus}=0.292$ (and were inhabited). The column header represents the numerator and the row header represents the denominator. The upper-panel assumes a Jaynes prior, the lower-panel assumes a uniform prior in $f$.
} 
\centering 
\begin{tabular}{c c c c c c} 
 & \vline & $\mathcal{M}_I$ & $\mathcal{M}_E$ & $\mathcal{M}_L$ & $\mathcal{M}_O$ \\ [0.5ex] 
\hline
$\mathcal{M}_I$ & \vline & 1 	& 2.74	& 0.34	& 2.01	\\
$\mathcal{M}_E$ & \vline & 0.37 	& 1		& 0.12	& 0.73	\\
$\mathcal{M}_L$ & \vline & 2.93 	& 8.02	& 1		& 5.88	\\
$\mathcal{M}_O$ & \vline & 0.50 	& 1.36	& 0.17	& 1		\\ [0.5ex]
\hline 
$\mathcal{M}_I$ & \vline & 1 	& 1.54	& 0.34	& 2.01	\\
$\mathcal{M}_E$ & \vline & 0.65 	& 1		& 0.22	& 1.30	\\
$\mathcal{M}_L$ & \vline & 2.93 	& 4.51	& 1		& 5.88	\\
$\mathcal{M}_O$ & \vline & 0.50 	& 0.77	& 0.17	& 1		\\ [0.5ex]
\hline
\end{tabular}
\label{tab:mars_matrix} 
\end{table}

\subsection{A Variable Second Datum}
\label{sub:variablemars}

As before, we can play the game of tweaking this second datum to create the desired result of $B>10$. In Table~\ref{tab:varmars_matrix}, we perform this task and find that a Martian data point could plausibly lead to a strong evidence outcome. The most moderate of these is the Bayes factor $B_{OL}$, for which we find that if i) Mars was inhabited, and ii) Mars had $f_{\mars}<0.20$, then the Earth and Mars together would represent strong evidence that inhabited planets follow an ocean-centric model versus a land-centric one. We argue such a result is at least plausible given the $f_{\mars}=0.25$ result of \citet{smith:1999}.

\begin{table}
\caption{
Matrix table evaluating how extreme Earth's land fraction would have needed to be in order to have led to Bayes factors exceeding $10$ (``strong'' evidence). The column header represents the numerator and the row header represents the denominator.  For example, the entry of $f_{\mars}<0.12$ corresponds to $B_{EL}$, the Bayes factor of the equi-centric (E) model over the land-centric (L) model. The upper-panel assumes a Jaynes prior, the lower-panel assumes a uniform prior in $f$. A ``-'' means no choice of $f$ within the interval $[0,1]$ provides strong evidence.
$\star$ indicates that $(1-f)$ greater than this value has the same effect.
} 
\centering 
\begin{tabular}{c c c c c c} 
 & \vline & $\mathcal{M}_I$ & $\mathcal{M}_E$ & $\mathcal{M}_L$ & $\mathcal{M}_O$ \\ [0.5ex] 
\hline
$\mathcal{M}_I$ & \vline & - & -	& -	& -	\\
$\mathcal{M}_E$ & \vline & $f_{\mars}<0.0076^{\star}$ & - & $f_{\mars}>0.99$	& $f_{\mars}<0.021$ \\
$\mathcal{M}_L$ & \vline & $f_{\mars}<0.086$ & $f_{\mars}<0.12$ & - & $f<0.20$ \\
$\mathcal{M}_O$ & \vline & $f_{\mars}>0.96$ 	& -	& $f_{\mars}>0.96$	& -		\\ [0.5ex]
\hline 
$\mathcal{M}_I$ & \vline & - & -	& -	& -	\\
$\mathcal{M}_E$ & \vline & $f_{\mars}<0.014^{\star}$ & - & $f_{\mars}>0.98$	& $f_{\mars}<0.038$ \\
$\mathcal{M}_L$ & \vline & $f_{\mars}<0.086$ & - & - & $f<0.20$ \\
$\mathcal{M}_O$ & \vline & $f_{\mars}>0.96$ 	& -	& $f_{\mars}>0.96$	& -		\\ [0.5ex]
\hline
\end{tabular}
\label{tab:varmars_matrix} 
\end{table}

\subsection{Adding Exoplanets}
\label{sub:exoplanets}

We note that, in principle, exoplanets could also one day represent new data points. Although testing between the land- and ocean-centric scenarios is useful, the intermediate equi-centric model is perhaps of greatest interest to test, in particular $B_{EO}$.

Using just the Earth, Table~\ref{tab:n1_matrix} reveals that $B_{EO}=1.17$ and adding in a hypothetical Mars vale of $f_{\mars}=f_{\oplus}$ in Table~\ref{tab:mars_matrix} boosts this to $B_{EO}=1.36$. Indeed, we find that this series simply follows $B_{EO}=1.168^x$ where $x$ is the number of Earth-clones added. Although we certainly don't expect all planets to resemble Earth, this at least provides a means of crudely estimating the sample size needed to reach strong evidence favouring $\mathcal{M}_E$ over $\mathcal{M}_L$. Evaluating, we find that 15 such planets (assuming a Jaynes prior, else this becomes 17 for a uniform prior) would be necessary including the Earth, and thus 13 or 14 exoplanets depending on our progress with Mars (or even Venus).

\section{Discussion}
\label{sec:discussion}

\subsection{Comparison to the previous work of Simpson (2017)}

\citet{simpson:2017} previously presented a Bayesian analysis of this problem and we take a moment to explain the differences between that study and this work. Fundamentally, the two works ask different questions. \citet{simpson:2017} seeks to infer the distribution of waterworlds given i) our existence upon the Earth, and, ii) a posited model for the observer selection effect. In contrast, we seek to compare different models for the observer selection effect, given i) our existence upon the Earth, and, ii) a maximally uninformative prior for the distribution of waterworlds. One is not superior to the other; they are simply asking different questions.

In our work, we consider that the observer selection effect takes four different forms: i) a land-centric ii) an ocean-centric iii) an equi-centric and iv) an indifferent model. \citet{simpson:2017} also includes a selection bias term but posit only a single model which takes the form

\begin{align}
\pdf(\mathrm{CES}|H,\mathcal{M}_{\mathrm{S17}}) &\propto H^2,
\label{eqn:likelihoodS}
\end{align}

where $H$ is a planet's habitable \textit{land} area. The quadratic scaling is the product of two linear effects: i) ``larger areas of habitable land permit a greater abundance and diversity of organisms to explore the evolutionary landscape'', and, ii) more land area means more observers. The model of \citet{simpson:2017} has a reasonable and well-motivated basis, but our approach here is much more agnostic and favours i) a more heuristic model formalism e.g. a power-law ii) a selection of such models, which we then compare using Bayesian model selection.

Our reasoning for this distinct approach is that there is considerable uncertainty about the assumptions implicit in Simpson's model. For example, one might reasonably question whether larger land areas genuinely lead to greater diversity, as testing that model would presumably require tracking the evolutionary processes on Earth-analogs with varying land fractions. Similarly, the second assumption treats land observers as the only relevant observers - why not ocean-dwelling ones?

\citet{simpson:2017} anticipated this criticism at the end of their Section~2.2, writing that there ``may be a number of water-based and land based observers, but a priori it is extremely unlikely that these two numbers are a similar order of magnitude'' and thus argue that our existence as land-observers essentially establishes the land-centric paradigm. Again, our work adopts a more agnostic view here, open to other observers beyond these two simple descriptions, such as amphibians, flying or lighter-than-air creatures, subterranean, etc. However, to be clear, \citet{simpson:2017} make a perfectly reasonable argument - but the difference here is that we elect to not treat any model as \textit{a-priori} true, rather we aim to test competing models based on the data. 

We also note that the prior ensemble of possible values for $f$ is quite different. \citet{simpson:2017} develop a physically-motivated model, considering basin saturation, water delivery and empirical mass-radius relations. We again make a much more simplistic choice of the maximally uninformative prior, on the basis that we know very little about the relative frequency or diversity of mechanisms driving $f$.

\subsection{Outlook}

Remarkably, even the single datum of our existence upon the Earth can be used to make useful inferences about life in the Universe. Indeed, we are not the first to squeeze this datum so; for example, \citet{simpson:2016} used the Earth's size to infer the largest habitable planet, \citet{kipping:2020} used life's early start to weigh the odds of rapid abiogenesis and \citet{loeb:2016} used the age of the Universe to explore its habitability over time.

The driving question behind this work is whether our existence upon a planet with nearly intermediate land fraction provides strong evidence that such planets are preferential when it comes to the emergence of observers such as ourselves. We find that simple bias models, linear in $f$, are not distinguishable to high confidence. Although this doesn't discredit the Rare Earth hypothesis in any way, it certainly robs it of ocean coverage bias as a statistically robust claim. However, more extreme models are already in tension with the our existence and we argue that future observational work, such as characterising other Earths, could firmly resolve the situation. Indeed, future observatories could plausibly detect oceans \citep{lustig:2018,ryan:2022} using glint effects, for example.



\section*{Acknowledgements}

Special thanks to donors to the Cool Worlds Lab, without whom this kind of research would not be possible:
Douglas Daughaday,
Elena West,
Tristan Zajonc,
Alex de Vaal,
Mark Elliott,
Stephen Lee,
Zachary Danielson,
Chad Souter,
Marcus Gillette,
Jason Rockett,
Tom Donkin,
Andrew Schoen,
Mike Hedlund,
Leigh Deacon,
Ryan Provost,
Nicholas De Haan,
Emerson Garland,
Queen Rd Fndn Inc,
Ieuan Williams,
Axel Nimmerjahn,
Brian Cartmell,
Guillaume Le Saint,
Robin Raszka,
Bas van Gaalen,
Josh Alley,
Drew Aron,
Warren Smith,
Brad Bueche,
Steve Larter,
Marisol Adler,
Craig Frederick,
Mathew Farabee \&
Philip Johnston.

\bibliography{sample701}{}
\bibliographystyle{aasjournalv7_noinit}



\end{document}